\documentclass[prb,twocolumn,showpacs,preprintnumbers,amsmath,amssymb,twoside]{revtex4}

\usepackage{graphicx}
\usepackage{dcolumn}
\usepackage{bm}
\usepackage[bookmarks={false}, pdfauthor={Rudro Rana Biswas}, pdftitle={Impurity-induced states on the surface of 3D topological insulators}]{hyperref}
\hypersetup{colorlinks=false, linkcolor=red, citecolor=green, filecolor=blue, urlcolor=blue, filebordercolor={.8 .8 1}, urlbordercolor={.8 .8 0}}
\usepackage[all]{hypcap}

\newcommand\ba{\begin{array}}
\newcommand\ea{\end{array}}
\newcommand\bp{\begin{picture}}
\newcommand\ep{\end{picture}}
\newcommand\be{\begin{equation}}
\newcommand\ee{\end{equation}}
\newcommand\bs{\begin{subequations}}
\newcommand\es{\end{subequations}}
\newcommand\nn{\nonumber}
\newcommand\bfl{\begin{flushleft}}
\newcommand\efl{\end{flushleft}}
\newcommand\bsp{\begin{split}}
\newcommand\easp{\end{split}}

\newcommand\ri{\right}
\renewcommand\le{\left}

\newcommand{\sub}[1]{\mbox{\tiny{#1}}}

\newcommand{\tto}{\rightarrow}
\newcommand{\tr}{\mbox{Tr}}
\newcommand{\sgn}{\mbox{sgn}}

\renewcommand\d{\delta}
\newcommand\D{\Delta}

\newcommand\m{\mu}

\newcommand\p{\pi}
\newcommand\mbp{\mbs{p}}

\newcommand\rr{\rho}
\newcommand\mbr{\mbs{r}}
\newcommand\mbrr{\mbs{\rr}}
\newcommand\s{\sigma}
\newcommand\mbss{\mbs{\s}}
\newcommand\mbsss{\mbs{s}}
\newcommand\mbS{\mbs{S}}

\renewcommand\th{\theta}

\newcommand\Th{\Theta}

\newcommand\w{\omega}
\newcommand\W{\Omega}

\newcommand\la{\langle}
\newcommand\ra{\rangle}

\newcommand\mc{\mathcal}
\newcommand\mb{\mathbb}
\newcommand\mbs{\boldsymbol}

\begin{document}

\title{Impurity-induced states on the surface of 3D topological insulators}
\author{Rudro R. Biswas$^{1,3}$}
\email{rrbiswas@physics.harvard.edu}
\author{A.  V.  Balatsky$^{2,3}$}%
\email{avb@lanl.gov}
\affiliation{%
$^{1}$Department of Physics, Harvard University, Cambridge, MA 02138\\
$^{2}$Theoretical Division, Los Alamos National Laboratory, Los Alamos, NM 87545\\
$^{3}$Center for Integrated Nanotechnologies, Los Alamos National Laboratory, Los Alamos, NM 87545
}
\date{\today}
\begin{abstract}
We calculate the modification of the local electronic structure caused by a single local impurity on the surface of a 3D Topological Insulator. We find that the LDOS around the Dirac point of the electronic spectrum at the surface is significantly disrupted near the impurity by the creation of low-energy resonance state(s) -- however, this is not sufficient to (locally) destroy the Dirac point. We also calculate the non-trivial spin textures created near the magnetic impurities and discover anisotropic RKKY coupling between them.
\end{abstract}
\maketitle

\section{Introduction}
\begin{figure*}
\resizebox{14cm}{!}{\includegraphics{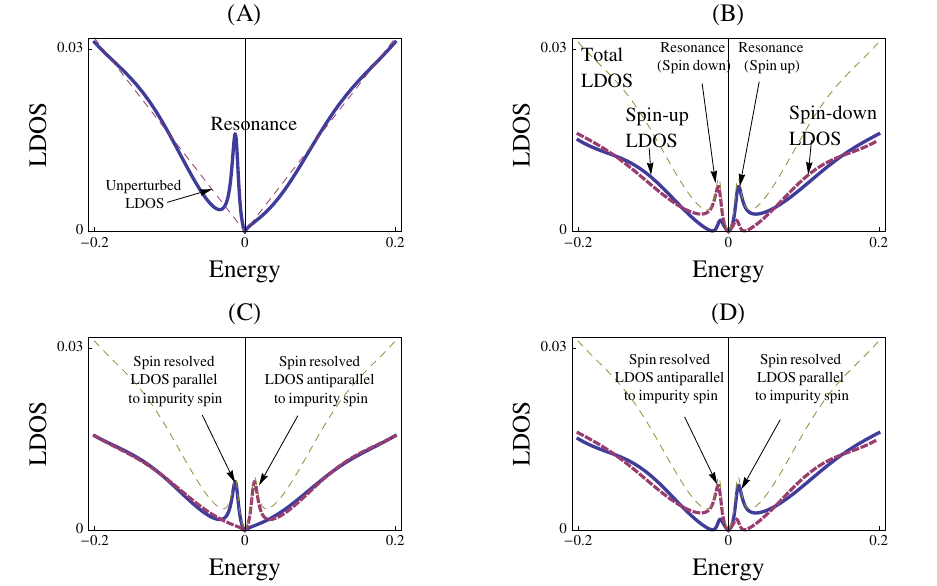}}
\caption{(Color Online) LDOS plots showing the low energy resonance(s) near (A) the scalar/potential impurity, (B) the $z$-polarized and (C,D) the $x$-polarized magnetic impurities. (C) and (D) show the $x$-spin projected LDOS, at a point on the $x$ and $y$-axis respectively. Note from (C) that on the $x$-axis, the negative energy states have excess states with spins \emph{parallel} to the $x$-polarized impurity. In all these cases, $U=100$, $r=20$. In the system of units used above, $\hbar$, $v_{F}$ and $W$ are unity.}
\label{fig-resonances}
\end{figure*}

The Dirac spectrum of chiral excitations are realized in a wide range of materials including d-wave superconductors\cite{2006-balatsky-yq}, graphene\cite{2009-neto-ys}, semiconductors\cite{1986-fradkin-yq} and superfluid $^{3}$He-A\cite{2003-volovik-fk}. The Dirac spectrum brings in substantial similarities in electronic properties -- like response to defects as well as low energy and low temperature properties. It is thus natural to combine these materials into a category of `Dirac materials'. A recent exciting realization of the Dirac spectrum is on the surface of 3D Strong Topological Insulators (STI)\cite{2009-xia-vn,2009-roushan-vn,2009-zhang-vn}. These materials have an ungapped spectrum at the surface while being fully gapped in the bulk. In addition, STIs are unique because the topology of their bulk band structure constrains their surface states to possess an odd number of Dirac nodes\cite{2007-fu-rt, 2008-schnyder-fk}. Suppressed backscattering inside the odd Dirac cone guarantees that the Dirac dispersion remain essentially unperturbed for any perturbation to the Hamiltonian that preserves time reversal symmetry. This is a manifestation of the topological protection enjoyed by this kind of surface band crossing and makes these materials an attractive candidate for spintronics applications\cite{2004-murakami-ul} as well as a possible platform for topological quantum computation\cite{2008-fu-yq}. In this context an important issue is the stability of the STI surface nodes to the presence of impurities\cite{2008-schnyder-fk,1994-nersesyan-uq,1985-gorkov-kx}. We contribute to this discussion by looking at the modification of surface states around a \emph{single} local potential/magnetic impurity and calculate the change in the Local Density of States (LDOS) as well as the spin density near the impurity site. These quantities should be accessible by STM measurements. We find the following.\\
(i) There is substantial modification of the LDOS near the impurity site for both the nonmagnetic (time reversal preserving) and magnetic impurities (time reversal breaking), especially when impurity scattering is strong (unitary). Near the potential/magnetic impurity, a single/a pair of low energy resonances form near the Dirac point (Figs. \ref{fig-resonances}, \ref{fig-ldos}). These become very sharp and their energies $\W\tto0$ as the impurity strength \eqref{eq-impurity} $|U|\tto\infty$ :
\begin{align}\label{eq-resscalar}
|\W| \approx \frac{5\,\sgn |U|}{|U| \ln |U|}
\end{align}
The scalar impurity resonance is doubly degenerate due to Kramers' theorem\cite{2007-wehling-td,2006-balatsky-yq,2005-bena-fr}. The magnetic impurity breaks time reversal symmetry and splits the low energy impurity resonance into two spin-polarized resonances on either side of the Dirac point (Fig. \ref{fig-resonances}).
\\
(ii) Modification of the LDOS vanishes quickly for energies less in magnitude than the resonance energy (approaching the Dirac point) for \emph{both} magnetic and nonmagnetic impurities. Thus, modifications to the low energy LDOS does not provide us with a signature of any incipient gap in the spectrum for both potential and magnetic impurities. For $r>>1/\w$, these decay as $1/r^{2}$.
\\
(iii) In addition to LDOS modifications, magnetic scattering produces non-trivial spin textures near the impurity site (Figure \ref{fig-spintextures}) that can be imaged with a magnetic force microscope or spin-resolved STM. These non-trivial spin textures lead to the propagation of unconventional {anti}ferromagnetic (AF) RKKY coupling between magnetic impurities, when they are polarized along the line joining them and when the chemical potential is close to the Dirac point. When the spins are perpendicular to the line joining them, they interact strongly and ferromagnetically (FM). The Dzyaloshinskii-Moriya (DM)  interaction between the spins\cite{2010-ye-fk} vanishes at the Dirac point. We thus conclude that random magnetic impurities will tend to align parallel to the normal to the STI surface.

\section{Theory}
We will model the STI surface states  as a single species of non-interacting 2-D Dirac quasiparticles\cite{2009-liu-ph} with a high energy band cutoff $W$. We shall work in units of $W$, $\hbar$ and $v_{F}$ (the Fermi velocity). The Hamiltonian becomes:
\begin{align}\label{eq-h0}
\mc{H}_{0} &= \mbs{\s}\cdot\mbp
\end{align}
where $\mbs{\s}/2$ is  the \emph{actual} spin of the electron (or related by a rotation about $\hat{z}$). We shall consider local impurities of the potential and classical types respectively:
\begin{align}\label{eq-impurity}
\hat{V}_{\sub{pot}} &= U\, \mb{I} \d(\hat{\mbs{r}}),\; \hat{V}_{\sub{mag}} = U \mbS\cdot\mbss\d(\hat{\mbs{r}})
\end{align}
For the magnetic case we have assumed a local Heisenberg exchange $J$ between the band electrons and the impurity spin $S$, whose direction is given by the \emph{unit vector} $\mbS$. Thus, $U=JS/2$ in $\hat{V}_{\sub{mag}}$.

To address the effect of impurity scattering we use the T-matrix technique \cite{2006-balatsky-yq}. The T-matrix is defined via:
\begin{align}
\hat{T}(\w) &= \hat{V} + \hat{V}\hat{G}_{0}^{\sub{ret}}(\w)\hat{T}(\w)
\end{align}
where $G_{0}^{\sub{ret}}$ is the retarded Green's function for the impurity-free material and $\w$ is the energy. For $\w\ll 1$ and $\rr\gg1/W$ ($\mbrr \equiv \mbr-\mbr'$), it has the following form:
\begin{align}\label{eq-propagator}
\le\la\mbr\le|G_{0}^{\sub{ret}}(\w)\ri|\mbr'\ri\ra &=\frac{|\omega |}{4} \le[f_{0}(\w, \rr)\mb{I} + f_{1}(\w,\rr)(\mbss\cdot\hat{\mbrr})\ri]
\end{align}
where
\begin{align}
f_{0}(\w, \rr) = \text{s}(\w)Y_{0} - i J_{0}\th, f_{1}(\w, \rr) = iY_{1} + \text{s}(\w) J_{1}\th
\end{align}
and $|\w|\rr$ is the argument of the Bessel functions $J_{0/1}$ and $Y_{0/1}$. Also, $\text{s}(\cdot)\equiv\sgn(\cdot)$ and $\th\equiv\Th(1 - |\w|)$. We shall also require the unperturbed on-site Green's function valid for short distances $\lesssim 1$:
\begin{align}\label{eq-onsiteG}
\mb{G}_{0}(\w) &\equiv \le\la\mbs{0}\le|G_{0}^{\sub{ret}}(\w)\ri|\mbs{0}\ri\ra = -\le(g_{0}(\w) + i g_{1}(\w)\ri)\mb{I},\text{ where}\nn\\
g_{0}(\w)&= \frac{\w}{4\p}\ln\le|\frac{1}{\w^{2}} - 1\ri|, g_{1}(\w) = \frac{|\w|}{4}\Th(1 - |\w|)
\end{align}
In \eqref{eq-impurity}, we have used a local form for the impurity potential $\la\mbr|\hat{V}|\mbr'\ra = \mb{V}\,\d(\mbr)\d(\mbr')$, where $\mb{V}$ is a $2\times2$ matrix in spin-space. The T-matrix also becomes $\la\mbr|\hat{T}|\mbr'\ra = \mb{T}\,\d(\mbr)\d(\mbr')$, with $\mb{T}$ satisfying the following equation
\begin{align}\label{eq-tmatrix2}
\mb{T} &= \mb{V} + \mb{V}\mb{G}_{0}\mb{T} = \le(\mb{I}-\mb{V}\mb{G}_{0}\ri)^{-1}\mb{V}
\end{align}
From the algebraic relations involving \eqref{eq-impurity}, \eqref{eq-onsiteG} and \eqref{eq-tmatrix2}, we analytically calculate the T-matrix, the full Green's function $\hat{G}^{\sub{ret}}$,
\begin{align}\label{eq-greens}
\hat{G}^{\sub{ret}}(\w) &= \hat{G}_{0}^{\sub{ret}}(\w) + \hat{G}_{0}^{\sub{ret}}(\w)\hat{T}(\w)\hat{G}_{0}^{\sub{ret}}(\w)
\end{align}
the full (spin-unresolved) LDOS,
\begin{align}
\rr(\mbr, \w) = -\frac{1}{\p}\text{Im}\tr\le\la\mbr\le|\hat{G}^{\sub{ret}}(\w)\ri|\mbr\ri\ra,
\end{align}
the local density of spin up/down states (in direction $\m$),
\begin{align}\label{eq-SPldos}
\rr_{\pm}^{\m}(\mbr, \w) = -\frac{1}{\p}\text{Im}\tr\le\la\mbr\le|\hat{G}^{\sub{ret}}(\w)\le(\frac{1\pm\s_{\m}}{2}\ri)\ri|\mbr\ri\ra
\end{align}
and the energy-resolved spin density averages:
\begin{align}
\mbs{s}(\mbr, \w) = -\frac{1}{\p}\text{Im}\tr\le\la\mbr\le|\hat{G}^{\sub{ret}}(\w)\frac{\mbs{\s}}{2}\ri|\mbr\ri\ra
\end{align}

\section{Results}
For the scalar and magnetic impurity cases, we find that the additional $GTG\equiv\d G$ pieces in the Green's function \eqref{eq-greens} evaluate respectively to (using $g \equiv g_{1}+i g_{2}$)
\begin{align}\label{eq-gtg1}
\d G_{\text{pot}} &= \frac{U\w^{2}}{16}\frac{f_{0}^{2} - f_{1}^{2}}{1+Ug}
\end{align}
and
\begin{align}\label{eq-gtg2}
&\d G_{\text{mag}} = \frac{U\w^{2}}{16(1-U^{2}g^{2})}\big[-2if_{0}f_{1}\mbss\cdot(\mbS\times\hat{\mbr})+\nn\\
&(f_{0}^{2} + f_{1}^{2})\mbss\cdot\mbS - 2f_{1}^{2}(\mbss\cdot\hat{\mbr})(\mbS\cdot\hat{\mbr})-Ug (f_{0}^{2} - f_{1}^{2}) \big]
\end{align}
\begin{figure}
\resizebox{7cm}{!}{\includegraphics{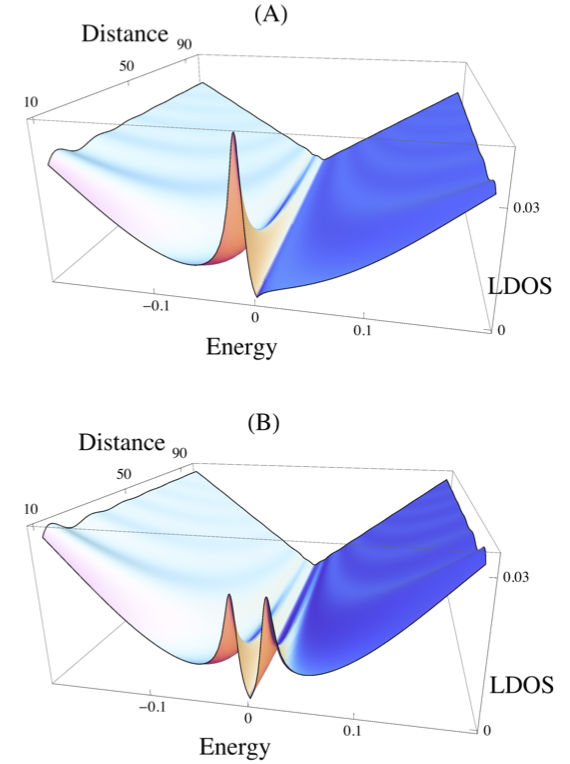}}
\caption{(Color Online) Low energy LDOS near the (A) scalar impurity and (B) the $x$-polarized magnetic impurity for $U=80$. $\hbar$, $v$ and $W$ have been set to unity.}
\label{fig-ldos}
\end{figure}
As shown in Figures (\ref{fig-resonances}) and (\ref{fig-ldos}), for both the magnetic and non-magnetic cases we obtain low energy resonance(s) in the LDOS (arising from the minima of the denominators in \eqref{eq-gtg1} and \eqref{eq-gtg2}) that approach the Dirac point for large impurity strengths according to \eqref{eq-resscalar} \cite{2007-wehling-td,2006-balatsky-yq}. These resonances become sharper as they approach the Dirac point with increasing potential strength and while doing so, also increase in amplitude relative to the unperturbed LDOS. For $r\gg1/\w$ and $\w\ll1$, the strength of LDOS modulations diminish with distance as $1/r^{2}$\cite{2008-bena-fk}.

Topological stability of the surface Dirac spectrum in TIs is often discussed as a crucial property of these materials. An important question in this context is whether the appearance of these low-energy resonances is related to the local creation of a gap at/destruction of the Dirac point. Na\"{i}ve scaling analysis tells us that the potential strength $U$ has a dimension of $-1$ (same as length) near the fixed point corresponding to \eqref{eq-h0}. As we approach the Dirac point, we should thus see the effects of the impurity become negligible. Indeed, we find that if we move from the resonances to the Dirac point, the density of states gradually settles down to the impurity-free value. We cannot, therefore, find signatures of gap-opening at the Dirac point at the stage of one-impurity scattering. We also note here that the appearance of these resonances \emph{at} the Dirac point is a consequence of the band cutoff being symmetric on the particle and hole sides -- in realistic materials\cite{2009-xia-vn} the band structure is asymmetric and depending on the degree of asymmetry, these resonances may appear at other region(s) of the bands\cite{2006-balatsky-yq}.

In addition to the impurity resonances at small energy we find new states that lie outside the effective band edges -- a consequence of using a hard cutoff. These true bound/anti-bound states are located at the positive/negative side for positive/negative sign of a scalar impurity potential $U$. For a magnetic impurity they are located on both sides outside the effective band edges. For large $|U|$, these are located approximately at a distance $U$ from the Dirac point, while as $|U|\tto0$ they approach the band edge as $e^{-4/|U|}$. In real STI SSs, these may well be located at the same energy as the bulk bands, will hybridize with them and delocalize into the bulk.

\begin{figure}
\resizebox{8.7cm}{!}{\includegraphics{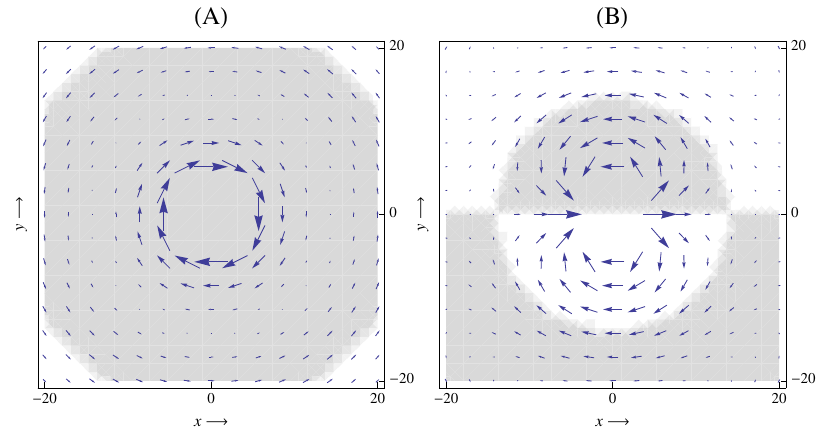}}
\caption{(Color Online) Spin textures near spin impurities ($U=80$, $E =\W= -0.014$) when the impurity is (A) $z$-polarized and when it is (B) $x$-polarized (solid green arrows). The component in the $xy$ plane is denoted by a vector while the background shade gives the sign of $s_{z}(r,E)$ (clear $\equiv$ positive). The arrows are normalized to the longest field-of-view total spin length in (A) and $xy$ spin length in (B), indicating respectively the sign of the $z$ polarization and the anisotropic $x$ polarization around the impurities in accordance with \eqref{eq-spinaverage} (these mediate \emph{anisotropic} RKKY interactions between two impurity spins).}
\label{fig-spintextures}
\end{figure}

Near a magnetic impurity, entanglement of the electron spin and momentum lead to the creation of spin textures, as shown in Figure~\ref{fig-spintextures}. The energy-resolved spin average is found to be:
\begin{align}\label{eq-spinaverage}
&\mbs{s}(\mbr, \w)\\
&= U\w^{2}\text{Im}\le(\frac{2if_{0}f_{1}\mbS\times\hat{\mbr} - (f_{0}^{2} + f_{1}^{2})\mbS + 2f_{1}^{2}\hat{\mbr}(\mbS\cdot\hat{\mbr})}{16\p(1-U^{2}g^{2})}\ri)\nn
\end{align}
The first term in \eqref{eq-spinaverage} gives rise to a DM interaction between two impurity spins. When the chemical potential $\m$ is at the Dirac point, considering only the perturbative result (obtained cheaply by putting $g\tto0$ in the above expression), the strength of this interaction becomes zero
\begin{align}
&\int_{-\infty}^{0}\text{Re}(f_{0}f_{1})\w^{2}d\w \nn\\
&\sim- \frac{1}{r^{3}} \int_{0}^{\infty}\frac{d}{dx}(J_{0}Y_{0})\Th(r-x)x^{2}dx \stackrel{r\gg1}{\approx} 0
\end{align}
For a finite chemical potential $|\m|\ll 1$, the amplitude of the DM interaction becomes
\begin{align}
&\frac{U}{8\p}\int_{-\infty}^{\m}\text{Re}(f_{0}f_{1})\w^{2}d\w= \frac{U\sgn\m}{8\p}\int_{0}^{|\m|}\text{Re}(f_{0}f_{1})\w^{2}d\w\nn\\
&= \frac{U\m|\m|J_{1}(|\m|r)Y_{1}(|\m|r)}{8\p r}
\end{align}
At large distances, the amplitude of this interaction decays as $\sim U\m/r^{2}$.

The second term in \eqref{eq-spinaverage} leads to FM RKKY interactions when $\m=0$, in the perturbative approximation. The corresponding spin component is
\begin{align}
-\frac{U\mbS}{16\p}\int_{-\infty}^{0}\text{Im}(f_{0}^{2}+f_{1}^{2})\w^{2}d\w = -\frac{U}{32\p r^{3}}\mbS
\end{align}

Finally, the third term in \eqref{eq-spinaverage} leads to AF RKKY interaction between impurity spin components pointing along the line joining the impurities, when $\m=0$. In the perturbative limit, the corresponding induced spin component is:
\begin{align}
\frac{U\hat{\mbr}(\mbS\cdot\hat{\mbr})}{8\p}\int_{-\infty}^{0}\text{Im}(f_{1}^{2})\w^{2}d\w =\frac{3U}{64\p r^{3}}\hat{\mbr}(\mbS\cdot\hat{\mbr})
\end{align}

From these two expressions we calculate that at $\m=0$ the interaction energy between two impurity spins $\mbS_{1,2}$:
\begin{align}
&\D E_{12}(\mbr_{21})_{\m=0}= \la U\mbS_1\cdot\mbsss_1 + U\mbS_2\cdot\mbsss_2\ra_{\text{cross terms}}\\
&=\frac{U^2}{16r^3}\le( -\,\mbS_1\cdot\mbS_2 + \frac{3}{2}\,(\mbS_1\cdot\hat{\mbr_{21}})(\mbS_2\cdot\hat{\mbr_{21}})\ri) + \mc{O}(U^4)\nn
\end{align}
 is \emph{minimized} when they are aligned parallel to each other and perpendicular to the line joining them. Thus, when many impurities are present (and $\m=0$), they will tend to point in the common direction where all gain the FM interaction energy -- along the $z$-direction, normal to the surface. This kind of FM ordering will be conducive to opening a gap in the STI surface state spectrum. Also, we note here that this state does not arise due to the spontaneous breaking of a continuous symmetry and hence is not forbidden by the Mermin-Wagner theorem\cite{1966-mermin-oq}.

We would like to note here that the foregoing results are not obvious when observing the energy-resolved spin densities at low energies $\w\tto0-$, because of the low density of states there. Na\"{i}vely, one would have expected the low energy long wavelength features to determine the $r\tto\infty$ spin textures, but the low energy spin textures predict, incorrectly, antiferromagnetic RKKY interactions with short distance ferromagnetic contributions arising from non-perturbative effects. We would also like to note here that we assumed a smooth cutoff when adding up the spin textures at different energies to eliminate cutoff-dependence\cite{2007-saremi-fk}. We have used multiplicative functions like $e^{-\eta|\w|}$ and $(1-e^{-\eta\w^{2}})/(\eta\w^{2})$ (having different characters as $|\w|\tto\infty$) in the energy integrals and then taken the limit $\eta\tto0+$ --- both these procedures gave the same limit\footnote{Calculations by the authors, on the lines of \cite{2009-shytov-yq}, have confirmed these results. These calculations will be presented in a later work.}.

When we consider the full nonperturbative spin average obtained by integrating \eqref{eq-spinaverage} numerically, the aforementioned behaviors seems to hold qualitatively if we look beyond the `ringing' introduced by a sharp cutoff.

\section{Conclusion}
In summary, we find that local impurities can strongly disrupt the structure near the Dirac node of 2-D surface states in 3-D topological insulators by forming low energy resonance(s). However, in the asymptotic approach to the Dirac point, the linear DOS is preserved, consistent with the negative scaling dimension of the impurity strength. Thus, the gap-opening mechanism for magnetic impurities is not evident at this stage of analysis. We also find that the induction of non-trivial spin textures near magnetic impurities leads to the mediation of antiferromagnetic RKKY coupling between impurity spin components parallel to the lines joining them, especially if the chemical potential is at the Dirac point (in which case the interaction does not oscillate in sign). The spin components perpendicular to the line joining the impurities, however, exhibit strong FM interaction. While there is, in general, a DM component in the spin interactions, it vanishes at the Dirac point.

\acknowledgements
We are grateful to D.\ Abanin, D.\ Basov, Z.\ Hasan, H.\ Manoharan,  N.\ Nagaosa, T. Wehling and Y.\ Xia for useful discussions -- especially to D.\ Abanin for drawing our attention to the method used in \cite{2009-shytov-yq} to calculate the RKKY interactions. This work was supported by the US DOE thorough BES and LDRD funding and by University of California UCOP program T027-09.

\end{document}